\documentclass[aps,amsmath,amssymb,prl,floatfix,reprint,superscriptaddress]{revtex4-2}
\usepackage{siunitx}
\usepackage{bm}
\usepackage{verbatim}
\usepackage{dcolumn}
\usepackage{graphicx}
\usepackage{hyperref}
\usepackage{listings}
\hypersetup{
	bookmarks=true,
	colorlinks=true,
	allcolors=blue
}
\usepackage{amsmath}

\newcommand{\jpsi}{J/\psi}
\newcommand{\piz}{\pi^{0}}

\newcommand*{\dif }{\mathop{}\!\mathrm{d}}

\newcommand{\SigRadDec} {\Sigma^+\to p\gamma }
\newcommand{\SigbRadDec}{\bar{\Sigma}^-\to \bar{p}\gamma}
\newcommand{\SigHadDec} {\Sigma^+\to p\piz}
\newcommand{\SigbHadDec}{\bar{\Sigma}^-\to \bar{p}\piz}
\newcommand{\Sig}{\Sigma^+}

\newcommand{\thep}{\theta_{p}}
\newcommand{\thepb}{\theta_{\bar{p}}}
\newcommand{\phip}{\phi_{p}}
\newcommand{\phipb}{\phi_{\bar{p}}}
\newcommand{\jpsiSS}{\jpsi\to\Sigma^+\bar{\Sigma}^-}
\newcommand{\jpsiSShh}{\jpsi\to\Sigma^+ \left( \rightarrow p\piz\right)\bar{\Sigma}^- \left( \rightarrow\bar{p}\piz\right)}
\newcommand{\jpsiSSrh}{\jpsi\to\Sigma^+ \left( \rightarrow p\gamma\right)\bar{\Sigma}^- \left( \rightarrow\bar{p}\piz\right)}

\newcommand{\ar}{\alpha_{\gamma}}

\newcommand{\ahb}{\bar{\alpha}_{0}}
\DeclareSIUnit\clight{\text{\ensuremath{c}}}

\begin{document}

\title{\boldmath Precision Measurement of the Decay $\Sigma^{+} \rightarrow
		p\gamma$ in the Process $J/\psi\rightarrow \Sigma^{+}\bar{\Sigma}^{-}$}

\author{
	M.~Ablikim$^{1}$, M.~N.~Achasov$^{13,b}$, P.~Adlarson$^{73}$, R.~Aliberti$^{34}$, A.~Amoroso$^{72A,72C}$, M.~R.~An$^{38}$, Q.~An$^{69,56}$, Y.~Bai$^{55}$, O.~Bakina$^{35}$, I.~Balossino$^{29A}$
% ...
	\\
	\vspace{0.2cm}
	(BESIII Collaboration)\\
	\vspace{0.2cm} {\it
		$^{1}$ Institute of High Energy Physics, Beijing 100049, People's Republic of China\\
% ...
$^{13}$ G.I. Budker Institute of Nuclear Physics SB RAS (BINP), Novosibirsk 630090, Russia\\
% ...
$^{29}$ INFN Sezione di Ferrara, (A)INFN Sezione di Ferrara, I-44122, Ferrara, Italy; (B)University of Ferrara,  I-44122, Ferrara, Italy\\
% ...
$^{34}$ Johannes Gutenberg University of Mainz, Johann-Joachim-Becher-Weg 45, D-55099 Mainz, Germany\\
		$^{35}$ Joint Institute for Nuclear Research, 141980 Dubna, Moscow region, Russia\\
% ...
$^{38}$ Liaoning Normal University, Dalian 116029, People's Republic of China\\
% ...
$^{55}$ Southeast University, Nanjing 211100, People's Republic of China\\
		$^{56}$ State Key Laboratory of Particle Detection and Electronics, Beijing 100049, Hefei 230026, People's Republic of China\\
% ...
$^{69}$ University of Science and Technology of China, Hefei 230026, People's Republic of China\\
% ...
$^{72}$ University of Turin and INFN, (A)University of Turin, I-10125, Turin, Italy; (B)University of Eastern Piedmont, I-15121, Alessandria, Italy; (C)INFN, I-10125, Turin, Italy\\
		$^{73}$ Uppsala University, Box 516, SE-75120 Uppsala, Sweden\\
% ...
$^{b}$ Also at the Novosibirsk State University, Novosibirsk, 630090, Russia\\
% ...
	}
}

\date{\today}

\begin{abstract}
	Using \num{10087 \pm 44 e6} $\jpsi$ events collected with the BESIII
	detector, the radiative hyperon decay $\SigRadDec$ is studied at an
	electron-positron collider experiment for the first time. The
	absolute branching fraction is measured to be $
		\left(\num[parse-numbers=false]{0.996 \pm 0.021_{\rm stat.}\pm
			0.018_{\rm syst.}}\right)\times 10^{-3}$, which is lower than its
	world average value by \num{4.2} standard deviations. Its decay
	asymmetry parameter is determined to be $\num[parse-numbers=false]{-0.652 \pm
			0.056_{\rm stat.}\pm 0.020_{\rm syst.}}$. The branching
	fraction and decay asymmetry parameter are the most precise to date, and the
	accuracies are improved by \SI{78}{\percent} and \SI{34}{\percent},
	respectively. 
	
\end{abstract}
\maketitle

Radiative hyperon decays provide valuable insight into the nature of
nonleptonic weak interactions~\cite{PhysRev.111.1691}. In general, the
radiative decays of a spin-$\frac{1}{2} $ hyperon are described by a parity conserving (P-wave) and a parity
violating (S-wave) amplitude. The non-vanishing parity
violating amplitude produces an asymmetric angular distribution of the daughter baryon in
the hyperon rest frame, as $\dif N/\dif\Omega=(N/4\pi)(1+\alpha_{\gamma}\mathbf{P}_i\cdot\hat{\mathrm{p}})$. Here $\mathbf{P}_i$ is the
polarization of the decaying hyperon, $\hat{\mathrm{p}}$ is the unit vector of the daughter baryon's momenta in the hyperon rest frame, and $\alpha_\gamma$
is the decay asymmetry parameter characterizing the mixing of S- and P-waves. In the limit of
unitary symmetry, the parity violating amplitude of radiative hyperon 
decays is predicted to be small, resulting in $\alpha_\gamma=0$~\cite{PhysRevLett.12.378}.  However, the
$\alpha_\gamma$ values measured in experiments are
large~\cite{PhysRevLett.59.868,PhysRevLett.63.2717,
	*PhysRevLett.64.843,*PhysRevLett.72.808,*PhysRevLett.86.3239,PhysRevLett.68.3004}.
Of all radiative hyperon decays, the $\SigRadDec$ decay was
observed first and stimulated continued controversy over
several decades.  It was first observed in bubble chamber
experiments~\cite{PhysRevLett.14.154,*PhysRev.188.2077,*MANZ1980217,*Ang:1969hg},
and was studied later at modern particle physics
spectrometers~\cite{PhysRevLett.59.868,Hessey:1989ep,*Bristol-Geneva-Heidelberg-Lausanne-QueenMaryColl-Rutherford:1985ksh}. The
current world average value~\cite{10.1093/ptep/ptaa104} is dominated
by precise measurements using a polarized charged hyperon beam at
Fermilab~\cite{PhysRevLett.68.3004,PhysRevD.51.4638}, where both
branching fraction (BF) and $\ar$ are obtained as ratios to
those of the $\SigHadDec$ decay.

Various phenomenological models have been proposed to explain the
experimental results of radiative hyperon
decays~\cite{gavela_parity_1981,nardulli_pole_1987,Niu_2020,balitsky_radiative_1989,*chang_challenges_2000,*dubovik_weak_2008,*Zenczykowski:2020hmg},
but none of them gives a unified picture in describing all radiative
hyperon decays. The theoretical predictions given by
Ref.~\cite{Shi:2022dhw} in the framework of chiral perturbation theory
(ChPT) are consistent with the experimental results of almost all
radiative hyperon decays except for $\SigRadDec$. Therefore, more
precise measurements on $\SigRadDec$ are crucial to test
theoretical approaches such as ChPT. In addition, it has
long been argued that $\Sigma^+\rightarrow pl^+l^- ~(l=e,\mu)$ decays
are good probes for physics beyond the standard model
(SM)~\cite{Pospelov:2008zw,Geng_20221,He_20181}. More precise BF and
$\alpha_\gamma$ measurements of $\SigRadDec$ offer critical
information on the form factor of $\Sigma^+\rightarrow
	pl^+l^-$~\cite{He_20181}, and thus provide better constraints on the SM
predictions for these decays. 

The unique properties of $\jpsiSS$ events produced in
electron-positron collisions, \emph{e.g.} the spin correlation and
transverse polarization of hyperon and
anti-hyperon~\cite{BESIII:2018cnd,BESIII:2020fqg}, and the well
defined kinematics, provide ideal conditions to study the decay
$\SigRadDec$.  A double-tagged technique is applied to determine the
absolute BF~\cite{PhysRevLett.56.2140}.  This approach was
applied in a recent work on the $\Lambda \rightarrow n\gamma$ decay
investigated in the $\jpsi \rightarrow \Lambda \bar{\Lambda}$ process
by \mbox{BESIII}~\cite{BESIII:2022rgl}, which achieved a better
precision than fixed target experiments. Moreover, the
decay $\SigRadDec$ and the corresponding charge conjugate one provide a good
opportunity to search for charge-parity ($CP$)
violation~\cite{Donoghue:1986hh,*Tandean:2002vy}, whose experimental
information is absent for
$\SigRadDec$~\cite{Li_1993,*PhysRevD.51.2271} currently.

In this Letter, using $\jpsiSS$ decays from \num{10087 \pm 44
	e6} $\jpsi$ events~\cite{BESIII:2021cxx} collected at \mbox{BESIII},
we report the measurements of the absolute BF and decay asymmetry parameter of
$\SigRadDec$, as well as a test of $CP$ violation in hyperon decays.
Throughout this Letter, charge conjugation is always
implied unless noted otherwise.

A detailed description of the design and performance of the
\mbox{BESIII} detector can be found in Ref.~\cite{ABLIKIM2010345,*HUANG2022142},
while the simulation and analysis software framework for \mbox{BESIII}
are described in Refs.~\cite{Li2006THEOS,*Asner:2009zza,*zhang2018low}.
First, events of $\jpsiSS$ are selected with a single-tag (ST)
approach, \emph{i.e.}~reconstructing a $\bar{\Sigma}^-$ candidate in one of
its dominant decay modes $\SigbHadDec$
($\mathrm{BF}=\SI{51.57\pm0.30}{\percent}$)~\cite{10.1093/ptep/ptaa104}.
Then, the double-tag (DT) decay $\SigRadDec$ is searched for in the
system recoiling against the ST $\bar{\Sigma}^-$ hyperon. The
corresponding absolute BF is calculated by
\begin{equation}
	\mathrm{BF}(\SigRadDec) = \frac{N^{\mathrm{obs}}_{\mathrm{DT}}~\varepsilon_{\mathrm{ST}}}{N^{\mathrm{obs}}_{\mathrm{ST}}~\varepsilon_{\mathrm{DT}}} ,
\end{equation}
where $N^{\mathrm{obs}}_{\mathrm{ST(DT)}}$ and $\varepsilon_{\mathrm{ST(DT)}}$ are the ST (DT) yields and the corresponding detection efficiencies,
respectively.

The helicity decay amplitudes of the
processes~\cite{FALDT201716,*PhysRevD.99.056008}, which will be used
for the measurement of $\ar$, are functions of the five observables
$\xi =
	(\theta_{\Sigma^+},\theta_p,\phi_p,\theta_{\bar{p}},\phi_{\bar{p}})$. Here,
$\theta_{\Sigma^+}$ is the angle between the $\Sig$ hyperon and the
electron beam in the $\jpsi$ rest frame, $\thep$ ($\thepb$) and
$\phip$ ($\phipb$) are the polar and azimuthal angles of the proton
(anti-proton) with respect to the $\Sig$ helicity frame,
respectively. The differential cross-section is given as
$\dif\sigma\propto\mathcal{W}(\xi)\dif\xi$ with
\begin{equation}
	\begin{aligned}
		\mathcal{W}(\mathcal{\xi}) = & \mathcal{F}_0(\xi)+\alpha_{\psi}\mathcal{F}_5(\xi)+\ar\bar{\alpha}_0                                                              \\
		                             & \times\left( \mathcal{F}_1(\xi)+\sqrt{1-\alpha^2_{\psi}}\cos(\Delta\Phi)\mathcal{F}_2(\xi)+\alpha_{\psi}\mathcal{F}_6(\xi)\right) \\
		                             & +\sqrt{1-\alpha_{\psi}^2}\sin(\Delta\Phi)(\ar \mathcal{F}_3(\xi)+\bar{\alpha}_0\mathcal{F}_4(\xi)),
	\end{aligned}
	\label{eq:dc2}
\end{equation}
where $\alpha_\psi$ and $\Delta\Phi$ are the hyperon production
parameters of the process $\jpsiSS$, $\bar{\alpha}_0$ is the decay asymmetry parameter for the $\SigbHadDec$ decays, and
$\mathcal{F}_i(\xi)$ ($i=0,1,...,6$) are the angular functions as described in detail in Ref.~\cite{FALDT201716}. The parameters used here are fixed to the values in Ref.~\cite{PhysRevLett.125.052004}, except for $\ar$ to be determined in this
analysis.

A sample of Monte Carlo (MC) simulated events of generic $\jpsi$ decays corresponding to the
luminosity of data is used to study possible background reactions. On
the ST side, the signal MC sample of $\jpsi \rightarrow\Sigma^+ \left(
	\rightarrow\mathrm{anything}\right)\bar{\Sigma}^- \left(
	\rightarrow\bar{p}\piz\right)$ is generated with its helicity decay
amplitude. A phase space (PHSP) MC sample of $\jpsi
	\rightarrow\Delta(1232)^+
	(\rightarrow{\mathrm{anything}})\bar{\Delta}(1232)^-(\rightarrow
	\bar{p}\piz)$ is generated to study ST background. On the DT side, MC
samples for the $\jpsiSSrh$ signal and the dominant background
$\jpsiSShh$ are generated according to their helicity decay
amplitudes.

Charged tracks and photon candidates are detected in the main drift
chamber and electromagnetic calorimeter (EMC), respectively. 
Requirements on the polar angle, distance of closest approach to the
interaction point, particle identification probability of charged
tracks, the polar angle, energy, EMC time and distance to the nearest
charged track of photon candidates are
applied as in Ref.~\cite{PhysRevLett.125.052004}. The $\piz$ candidates are
reconstructed from pairs of photon candidates with an invariant mass
$M_{\gamma\gamma}$ satisfying $\SI{116}{MeV/\clight^2} <
	M_{\gamma\gamma} < \SI{148}{MeV/\clight^2}$. A kinematic fit
constraining their invariant mass to the $\piz$
mass~\cite{10.1093/ptep/ptaa104} is performed on these photon pairs,
and the updated four-momentum is used in further analysis.

The candidate ST $\bar{\Sigma}^-$ events are required to have at least
one $\bar{p}$ and one $\piz$, and the $\bar{p}\piz$ combination is
required to have an invariant mass of
$\left|M_{\bar{p}\piz}-M_{\bar{\Sigma}^-}\right|<
	\SI{13.5}{MeV/\clight^2}$, where $M_{\bar{\Sigma}^-}$ is the
nominal $\bar{\Sigma}^-$ mass~\cite{10.1093/ptep/ptaa104}. If there is
more than one combination satisfying this requirement, the one with
the minimum $\left|M_{\bar{p}\piz}-M_{\bar{\Sigma}^-}\right|$ is kept
for further analysis.  Finally, the ST yield is obtained from the mass
distribution recoiling against the reconstructed $\bar{\Sigma}^-$
candidate, which is defined as
\begin{equation}
	M_{\mathrm{rec}}=\sqrt{ \left( E_{\mathrm{cms}}-E_{\bar{p}}-E_{\piz}\right)^2/c^4- \left( \mathbf{P}_{\bar{p}}+\mathbf{P}_{\piz}\right)^2/c^2 }.
\end{equation}
Here, $E_{\mathrm{cms}}$ is the center-of-mass energy, $E_{\bar{p}}$
($\mathbf{P}_{\bar{p}}$) and $E_{\piz}$ ($\mathbf{P}_{\piz}$) are the
energies (momenta) of the anti-proton and $\piz$ in the $\jpsi$ rest frame, respectively.

\begin{figure}[htbp]
	\includegraphics[width=0.8\linewidth]{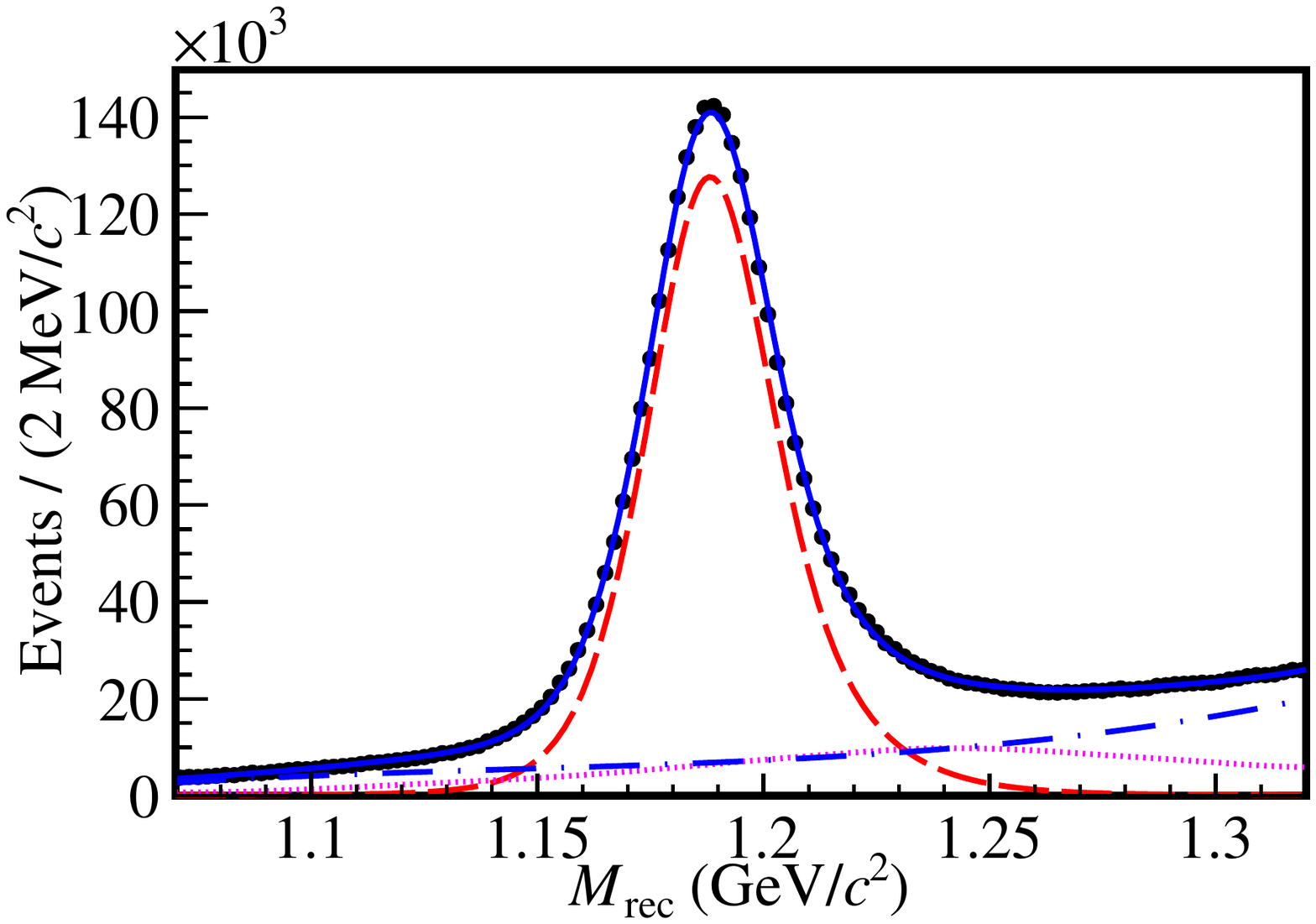}
	\caption{Fit to the $M_{\rm{rec}}$ distribution of ST
		candidates in data. Points with error bars represent
		data. The blue solid curve is the total fit result. The red
		dashed curve, magenta dotted curve and blue dash-dotted
		curve show the shapes of signal, $\jpsi
			\rightarrow\Delta(1232)^+ (\rightarrow{\mathrm{anything}}
			)\bar{\Delta}(1232)^-(\rightarrow \bar{p}\piz)$ background
		and other non-peaking background components, respectively.}%
	\label{fig:data_st_1}
\end{figure}

With the above selection criteria, the distribution of $M_{\rm{rec}}$
of the ST candidates is shown in Fig.~\ref{fig:data_st_1}, where there
is a clear peak representing the signal of $\jpsiSS$.  For ST
candidates, the background events are dominated by mis-combinations
and $\jpsi \rightarrow\Delta(1232)^+
	(\rightarrow{\mathrm{anything}})\bar{\Delta}(1232)^-(\rightarrow
	\bar{p}\piz)$, which contributes as a broad peak.  An unbinned maximum
likelihood fit is performed on the $M_{\rm{rec}}$ distribution to
determine the ST yield, with the fit result shown in
Fig.~\ref{fig:data_st_1}.  In the fit, the signal and $\jpsi
	\rightarrow\Delta(1232)^+
	(\rightarrow{\mathrm{anything}})\bar{\Delta}(1232)^-(\rightarrow
	\bar{p}\piz)$ background are described by their MC simulated shapes
convolved with a Gaussian function, which represents the resolution
difference between data and MC simulation. Other non-peaking
background is described by a third order polynomial
function.  The ST yield and the detection efficiency evaluated using
the signal MC sample are summarized in Table~\ref{tab:br}. 

The signal process $\SigRadDec$ is searched for in the remaining
charged tracks and photons recoiling against the selected ST
$\bar{\Sigma}^-$ candidates.  One remaining $p$ and at least one
remaining photon candidate are required for DT candidate events.  A
five-constraint (5C) kinematic fit under the hypothesis $\jpsi
	\rightarrow p\bar{p}\piz\gamma$ is performed. The fit constrains
the total energy-momentum of the final state particles to the four-momentum of $\jpsi$ and the invariant mass of the photon pair
on the ST side to the $\piz$ nominal
mass~\cite{10.1093/ptep/ptaa104}. The $\chi^2$ of the 5C kinematic fit
($\chi^2_{\mathrm{5C}}$) is required to be less than \num{30}.  In
case of multiple photon candidates on the DT side, the one with the
minimum $\chi^2_{\mathrm{5C}}$ is kept. This requirement removes $\sim$\SI{68}{\percent} of the background at the cost of \SI{13.9}{\percent} of the
signal.

MC studies with a generic event type analysis tool~\cite{zhouxy} indicate that the dominant background events on the DT side
are from the processes $\jpsi \rightarrow \Sigma^+ (\rightarrow
	p\piz)\bar{\Sigma}^-(\rightarrow \bar{p}\piz)$ and $\jpsi \rightarrow
	\Delta(1232)^+(\rightarrow p\piz)\bar{\Delta}(1232)^-(\rightarrow
	\bar{p}\piz)$, denoted as $\Sigma^+\rightarrow p\piz$ and
$\Delta(1232)^+\rightarrow p\piz$ background. To remove the
$\Delta(1232)^+\rightarrow p\piz$ background, we make use of the life
time difference between $\Sigma^+$ and $\Delta(1232)^+$.  A secondary
vertex fit~\cite{Min_2010} is performed for the $p$ and $\bar{p}$ combination. The length ($L$) from the intersection point to
the interaction point shows significantly different distribution for the $\Delta(1232)^+\rightarrow p\piz$ background and the
signal, as shown in Section~2 of the supplemental material~\cite{BESIII:Supplemental}. The $L/\sigma_L$ value of the $\Delta(1232)^+\rightarrow p\piz$
background is generally less than \num{1.5}, where
$\sigma_L$ is the decay length resolution. By vetoing these events, the $\Delta(1232)^+\rightarrow p\piz$ background is reduced by \SI{93}{\percent} while
\SI{78}{\percent} of the signal is preserved. The remaining $\Sigma^+
	\rightarrow p\piz$ background usually includes a high energy photon
from an asymmetric $\piz$ decay. To suppress this background, a 5C
kinematic fit under the $\jpsi\to p\bar{p}\piz\gamma\gamma$ hypothesis
is performed for all photon pair combinations. A candidate is
eliminated if any $\chi^2$ of the kinematic fit under the $\jpsi\to
	p\bar{p}\piz\gamma\gamma$ hypothesis is less than
$\chi^2_{5C}$ under the signal hypothesis.

\begin{figure}[htpb] \centering
	\includegraphics[width=\linewidth]{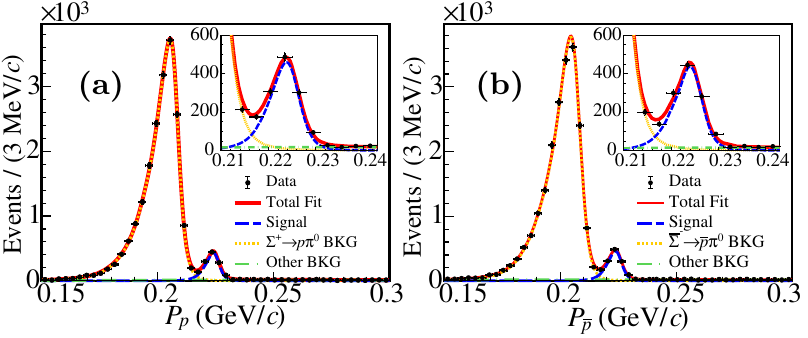}
	\caption{Fits to the $P_p$ distributions for (a)
		$\Sigma^+\rightarrow p\gamma$ and (b)
		$\bar{\Sigma}^-\rightarrow \bar{p}\gamma $ candidate
		events. Points with error bars are data. The red solid curves
		are the total fit results. The blue dashed, orange dotted and
		green dash-dotted curves show the shapes of signal,
		$\SigHadDec$ background and other background,
		respectively. The insets show the details of the fit in the
		signal region.}%
	\label{fig:fit_dt}
\end{figure}

After applying all the above selection criteria, the distribution of
the proton momentum in the rest frame of $\Sigma^+$ ($P_p$) is shown
in Fig.~\ref{fig:fit_dt}. The prominent peak at 0.204~GeV/$c$ is the
major background from $\SigHadDec$, and the peaking structure around
0.223~GeV/$c$ is from the $\SigRadDec$ signal. The contribution
from the $\Delta(1232)^+\rightarrow p\piz$ background is negligible.

To obtain the DT signal yield, an unbinned maximum likelihood fit is
performed on the $P_p$ distribution. In the fit, the $\SigRadDec$
signal and the $\SigHadDec$ background are described by MC
simulated shapes convolved with a Gaussian function. A second order
polynomial function is included to represent the residual
background such as $\Delta(1232)^+\rightarrow p\piz$.  The fit results
are shown in Fig.~\ref{fig:fit_dt},
and the DT signal yields are summarized in Table~\ref{tab:br}. In the fit, the yields of the $\SigHadDec$ and $\SigbHadDec$ background are \num{18227.7\pm135.0}
and \num{20334.9\pm142.6}, respectively. To
account for the small difference in the selection efficiencies of
charged tracks and photons between data and MC simulation, an
efficiency correction has been performed on the DT
side~\cite{PhysRevLett.125.052004}.
The resultant DT efficiencies and
the BFs are also summarized in Table~\ref{tab:br}.  The BFs for
$\SigRadDec$ and $\SigbRadDec$ are consistent with each other
within their uncertainties. Therefore, a simultaneous fit assuming the
same BF between the charge conjugate channels is performed. The
resultant BF is also shown in Table~\ref{tab:br}.

The decay asymmetry parameter $\ar$ is determined by an angular analysis
according to Eq.~(\ref{eq:dc2}).  To further improve the purity of the
data sample used for the $\ar$ measurements, two additional kinematic fits
under the hypotheses of $\jpsi\to p\bar{p}\piz+ \piz$ and $\jpsi\to
	p\bar{p}\piz+\gamma$ are performed for the surviving DT candidates,
where the momenta information of the $\piz$ and photon are missing in the fit. The DT
candidates are removed if $\chi^2_{p\bar{p}\piz+\piz}<\chi^2_{p\bar{p}\piz+\gamma}$, where
$\chi^2_{p\bar{p}\piz+\piz}$ and $\chi^2_{p\bar{p}\piz+\gamma}$ are the $\chi^2$ from the
two kinematic fits.  Finally, \num{2345} events
(including \num{243} background events) for the two charge conjugate
modes in the signal region $0.215 < P_p < \SI{0.235}{GeV/\clight}$ are
used in the following analysis.

$\ar$ is obtained from an unbinned maximum likelihood fit with the
likelihood function defined as 

\begin{equation} 
	\mathcal{L} \left( H\right)= \prod^{N}_{i=1} \frac{\mathcal{W} \left(
		\xi_i,H\right)\varepsilon \left( \xi_i\right)}{C \left(
		H\right)}, 
\end{equation}
where $H=\left(\alpha_{\psi},\Delta\Phi,\ar,\ahb\right)$ represents
the decay parameters, $N$ is the total number of DT candidate events,
$i$ is the corresponding event index, $\mathcal{W} \left(
	\xi_i,H\right)$ is the square of the decay amplitude as defined in
Eq.~(\ref{eq:dc2}), $\varepsilon \left( \xi_i\right)$ is the detection
efficiency and $C \left( H\right)$ is the normalization factor
calculated with the PHSP MC sample~\cite{BESIII:2018cnd}.  Due to the small
number of events, the parameters $\alpha_{\psi}$, $\Delta\Phi$ and $\ahb$ are fixed to
those in Ref.~\cite{PhysRevLett.125.052004} in the fit.

The value of $\ar$ is obtained by minimizing the likelihood function
$S=- \left(
	\ln\mathcal{L}_{\rm{data}}-\ln\mathcal{L}_{\rm{bkg}}\right)$, where
$\mathcal{L}_{\rm{data}}$ is the likelihood value from
the data sample, and $\mathcal{L}_{\rm{bkg}}$ represents the
background contributions including $\SigHadDec$ and other background
components.  The former background contribution is estimated with a
MC simulated event sample that is five times the size of data and is
generated according to its decay amplitude.  The latter is estimated
with data events in the side-band region defined as
\SI{0.11}{GeV/\clight} $<P_p<$ \SI{0.16}{GeV/\clight} and
\SI{0.24}{GeV/\clight} $<P_p<$\SI{0.29}{GeV/\clight}.
$\mathcal{L}_{\rm{bkg}}$ is estimated with the above samples and
normalized to data.  The fits are performed for the $\SigRadDec$ and
$\SigbRadDec$ decays individually. The $\ar$ results are
consistent within their uncertainties and summarized in
Table~\ref{tab:br}.  Further, a simultaneous fit, assuming the
same magnitude but opposite sign for the decay asymmetry parameters of the
charge conjugate channels, is also performed, and the result is also
shown in Table~\ref{tab:br}.

\begingroup
\squeezetable
\begin{table}[htbp]
	\caption{\small The values of $N_{\rm{ST}}^{\rm{obs}}$,
		$\varepsilon_{\rm{ST}}$, $N_{\rm{DT}}^{\rm{obs}}$, and
		$\varepsilon_{\rm{DT}}$ for the decays $\SigRadDec$
		and $\SigbRadDec$. The BF and $\alpha_{\gamma}$ are obtained from both individual and simultaneous fits. The first uncertainties are statistical and
		the second systematic.}
	\label{tab:br}
	\begin{ruledtabular}
		\begin{tabular}{cD{,}{\pm}{-1}D{,}{\pm}{-1}}
			
			Mode                                        & \multicolumn{1}{c}{$\Sigma^+\rightarrow p\gamma$}                     & 
			\multicolumn{1}{c}{$\bar{\Sigma}^-\rightarrow\bar{p}\gamma$}                                                                                   \\
			\colrule
			$N^{\mathrm{obs}}_{\mathrm{ST}}$            & \num{2177771},\num{2285}                                              & \num{2509380},\num{2301} \\
			$\varepsilon_{\mathrm{ST}}$ (\si{\percent}) & \num{39.00},\num{0.04}                                                & 
			\num{44.31},\num{0.04}                                                                                                                         \\
			$N^{\mathrm{obs}}_{\mathrm{DT}}$            & \num{1189},\num{38}                                                   & \num{1306},\num{39}      \\
			$\varepsilon_{\mathrm{DT}}$ (\si{\percent}) & \num{21.16},\num{0.03}                                                & \num{23.20},\num{0.03}   \\
			Individual BF (\num{e-3})                   & \num{1.005},\num{0.032}                                               & 
			\num{0.993},\num{0.030}                                                                                                                        \\
			Simultaneous BF (\num{e-3})                 & \multicolumn{2}{c}{\num[parse-numbers=false]{0.996\pm0.021\pm0.018}}                             \\
			\colrule
			Individual $\ar$                            & \num{-0.587},\num{0.082}                                              & 
			\num{0.710},\num{0.076}                                                                                                                        \\
			Simultaneous $\ar$                          & \multicolumn{2}{c}{\num[parse-numbers=false]{-0.651\pm0.056\pm0.020}}                            \\
		\end{tabular}
		
	\end{ruledtabular}
\end{table}
\endgroup

\begin{figure}[htbp]
	\includegraphics[width=\linewidth]{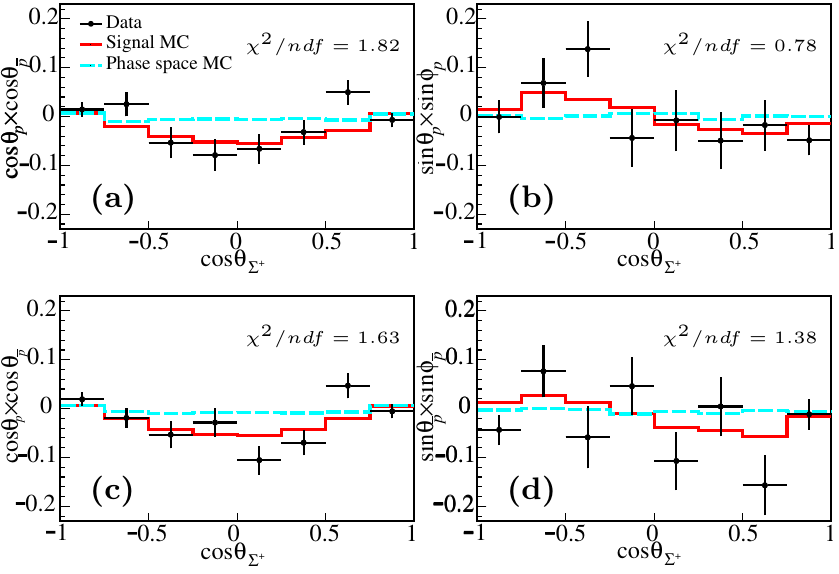}
	\caption{Moment distributions for (a)(b)
		$\bar{\Sigma}^-\rightarrow \bar{p}\gamma$ and (c)(d)
		$\SigRadDec$. The black points with error bars are
		data. The cyan dashed curves are the
		phase space MC simulation. The red solid curves show the MC simulation based on the fit results.
		For the comparison between charge conjugate channels,
		plots (c)(d) adopt the polar angle of $\bar{\Sigma}^-$
		hyperon as the abscissa variable.}%
	\label{fig:m12}
\end{figure}

To visualize the effect of the decay asymmetry, two moments are
calculated for $m=8$ intervals in $\cos\theta_{\Sigma^+}$:
\begin{equation}
	\begin{aligned}
		M_1(\cos\theta_{\Sigma^+})= \frac{m}{N} \sum^{N_k}_{i=1} \cos\theta_{\bar{p}}^{i}\cos\theta_{p}^{i}, \\
		M_2(\cos\theta_{\Sigma^+})= \frac{m}{N} \sum^{N_k}_{i=1} \sin \theta_{p}^{i}\sin \phi_{p}^{i},
	\end{aligned}
	\label{eq:mom} \end{equation} where $N$ is the total number of
events and $N_k$ is the number of events in the \textit{k}th
$\cos\theta_{\Sigma^+}$ interval. Figure~\ref{fig:m12} shows the
comparison of moments between data and MC projection based on
the fit. Good data-MC consistencies are observed.

The systematic uncertainties due to proton tracking and particle
identification (\SI{0.4}{\percent}), as well as photon detection
(\SI{0.3}{\percent}), are studied with $\jpsi \rightarrow
	p\bar{p}\pi^+\pi^-$ and $\jpsi \rightarrow \gamma\mu^+\mu^-$ control
samples. The uncertainties of $\chi^2$ requirements on the kinematic
fits (\SI{0.9}{\percent}) and decay length requirement
(\SI{0.4}{\percent}) are obtained by applying corresponding
requirements on a $\jpsiSShh$ control sample. The uncertainties of ST
(\SI{0.4}{\percent}) and DT (\SI{1.2}{\percent}) yields are estimated
by changing the fit parameters. The decay parameters used for signal
MC generation are varied by $\pm1\sigma$~\cite{PhysRevLett.125.052004}
to study the MC model related uncertainty
(\SI{0.6}{\percent}). Details can be found in the supplemental
material~\cite{BESIII:Supplemental}. By assuming all the sources to be
independent, the total systematic uncertainty in the BF measurement is
\SI{1.8}{\percent} by summing up all above values in
quadrature.

The systematic uncertainties in the determination of the decay
asymmetry parameter $\ar$ are separated into two categories: the fit-related
uncertainties and event selection uncertainties.  The fit-related
uncertainties are estimated with alternative fits by shifting the
sideband region (\num{0.004}), changing the signal region in the fit
(\num{0.014}), varying the number of background events by $\pm1\sigma$
(\num{0.002}) and varying the fixed decay parameters by
$\pm1\sigma$~\cite{PhysRevLett.125.052004} (\num{0.011}),
individually.

For the event selection induced uncertainties, only the angular
dependent event selection criteria are considered, since other effects
are negligible.  For each requirement, its dependence on the angular
distribution, and effect on the systematic uncertainty are detailed in
the supplemental material~\cite{BESIII:Supplemental}. The systematic
uncertainty due to the
track detection efficency (\num{0.001}) is estimated by performing
an efficiency correction on the PHSP MC. The decay length
requirement (\num{0.005}) and $\chi^2_{\gamma}<\chi^2_{\piz}$
requirement (\num{0.006}) uncertainties are studied with a
$\jpsiSShh$ control sample.  The total systematic uncertainty is
assigned to be \num{0.020} by adding all individual uncertainties
quadratically.

Based on the above results, the $CP$ asymmetry is calculated using the
BFs and decay asymmetry parameters between the charge conjugate channels:
\begin{equation*}
	\begin{aligned}
		 & \Delta_{CP}=
		\frac{\mathcal{B}_{+}-\mathcal{B}_{-}}{\mathcal{B}_{+}+\mathcal{B}_{-}}=\num[parse-numbers=false]{0.006\pm0.011_{\rm{stat.}}\pm0.004_{\rm{syst.}}}, \\
		 & A_{CP}= \frac{\alpha_-+\alpha_+}{\alpha_--\alpha_+}=\num[parse-numbers=false]{0.095\pm0.087_{\rm{stat.}}\pm0.018_{\rm{syst.}}}.
	\end{aligned} \end{equation*} Here, $\mathcal{B}_+$
($\alpha_+$) denotes the BF ($\ar$) of $\SigRadDec$ and
$\mathcal{B}_-$ ($\alpha_-$) is that of $\SigbRadDec$. The systematic uncertainties of $\Delta_{CP}$ and $A_{CP}$ only consider the uncorrelated
uncertainties
of $B_{+/-}$ and $\alpha_{+/-}$. This represents
the first search for $CP$ violation in $\SigRadDec$. No
obvious $CP$ violation is observed.

In summary, using \num{10087\pm44e6} $\jpsi$ events collected with the
\mbox{BESIII} detector, the radiative hyperon decay $\SigRadDec$ is
studied at an electron-positron collider for the first time.  The
absolute BF of this decay is determined to be $ (\num[tight-spacing =
		true, parse-numbers=false]{0.996 \pm 0.021_{\rm stat.}\pm 0.018_{\rm
				syst.}})\times 10^{-3}$, with a decay asymmetry parameter of
$\num[tight-spacing = true, parse-numbers=false]{-0.651 \pm 0.056_{\rm
				stat.}\pm 0.020_{\rm syst.}}$. Two independent observables between
the two charge conjugate channels, $A_{CP}$ and $\Delta_{CP}$, are
used to search for $CP$ violation, and no evidence of $CP$ violation
is found.
\begin{figure}[htpb]
	\centering \includegraphics[width=0.9\linewidth]{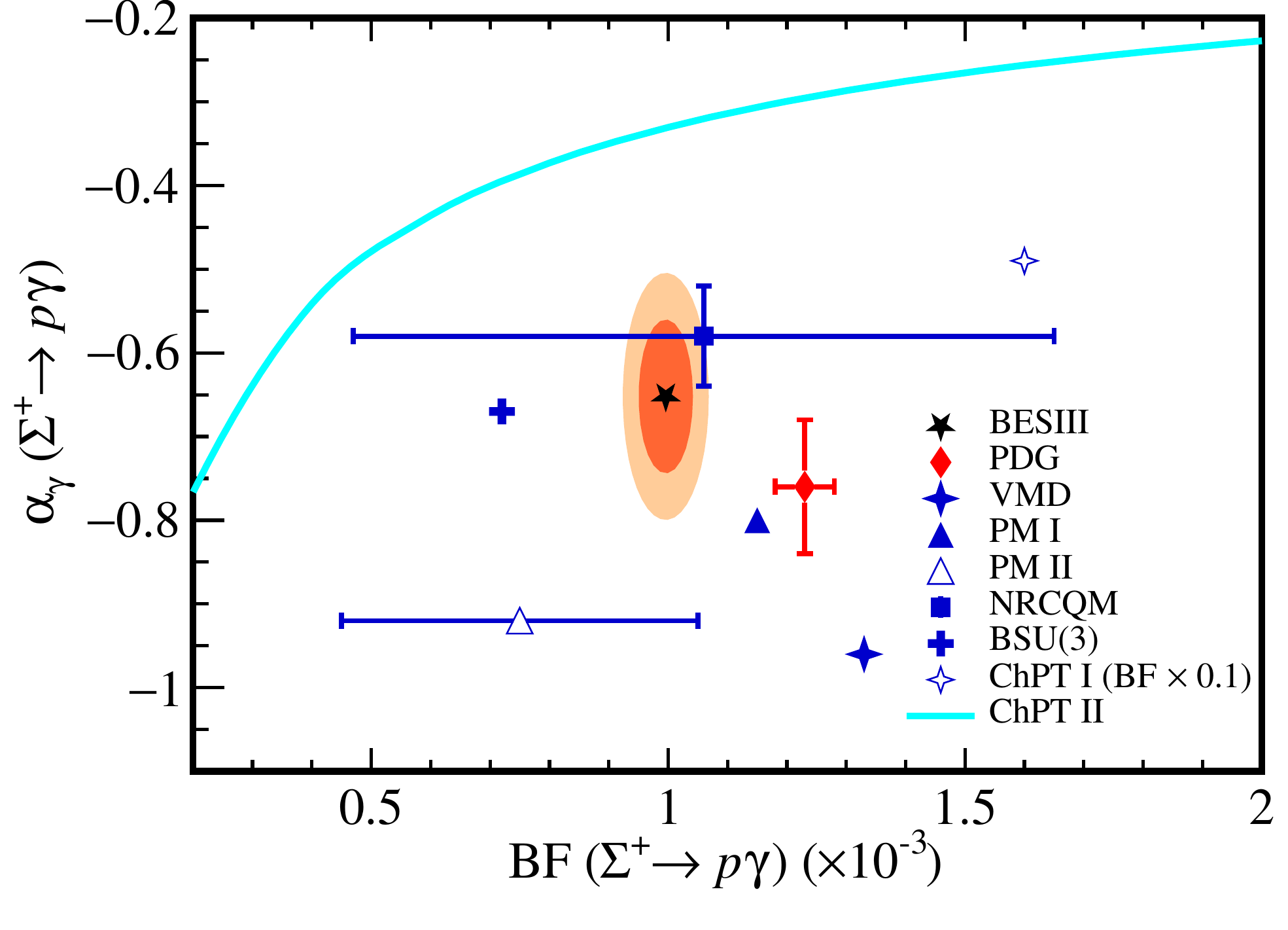}
	\caption{Distribution of $\alpha_{\gamma}$ versus BF of the
		$\SigRadDec$ decay. The black star denotes the results measured by
		this work and the orange contours correspond to the
		$\SI{68}{\percent}/\SI{95}{\percent}$ confidence-level of the
		results. The red diamond represents the PDG
		values~\cite{10.1093/ptep/ptaa104} of the BF and
		$\alpha_\gamma$. The cyan colored line shows the predicted $\alpha_{\gamma}$ as a function of BF cited from Ref.~\cite{Shi:2022dhw}. Other symbols in blue stand for the results
		predicted by the vector meson dominance model
		(VMD)~\cite{zenczykowski_reanalysis_1991}, pole model (PM I refers
		to Ref.~\cite{gavela_parity_1981} and PM II refers to
		Ref.~\cite{nardulli_pole_1987}), nonrelativistic constituent quark
		model (NRCQM)~\cite{Niu_2020}, broken SU(3) model
		(BSU(3))~\cite{zenczykowski_joint_2006} and
		another ChPT model~\cite{borasoy_resonances_1999}.
	}%
	\label{fig:result_com}
\end{figure}
The accuracies of the BF and $\ar$ are
improved by \SI{78}{\percent} and \SI{34}{\percent}, respectively. The comparison between the measurement results with the PDG values and
theoretical predictions are shown in Fig.~\ref{fig:result_com}. The
measured BF is lower than the PDG value~\cite{10.1093/ptep/ptaa104} by
$4.2\sigma$, where all previous measurement results were obtained as
relative ratios to $\SigHadDec$. The decay asymmetry parameter is consistent
with the world average value within $1.1\sigma$.  A similar
BF difference is also observed in another radiative hyperon decay
$\Lambda\to n\gamma$~\cite{BESIII:2022rgl} at BESIII.  With the
updated BF and $\alpha_\gamma$ results of $\SigRadDec$, further
theoretical efforts are needed to clarify the physics of radiative
hyperon decays.  The SM prediction of $\Sigma^+\rightarrow pl^+l^-$
decays can also be further improved~\cite{He_20181} to search for new
physics beyond the SM.

\begin{acknowledgments}
	The authors thank Prof. L. S. Geng and Prof. X. G. He for helpful discussion. The BESIII Collaboration thanks the staff of BEPCII and the IHEP computing center and the
	supercomputing center of USTC for their strong support. This work is supported in part by National Key R\&D Program of China under Contracts Nos.
	2020YFA0406300, 2020YFA0406400; National Natural Science Foundation of China (NSFC) under Contracts Nos. 11335008, 11625523, 11635010, 11705192,
	11735014, 11835012, 11935015, 11935016, 11935018, 11950410506, 11961141012, 12022510, 12025502, 12035009, 12035013, 12061131003, 12105276, 12122509,
	12192260, 12192261, 12192262, 12192263, 12192264, 12192265; the Chinese Academy of Sciences (CAS) Large-Scale Scientific Facility Program; the CAS
	Center for Excellence in Particle Physics (CCEPP); Joint Large-Scale Scientific Facility Funds of the NSFC and CAS under Contract No. U1732263,
	U1832103, U1832207, U2032111; CAS Key Research Program of Frontier Sciences under Contracts Nos. QYZDJ-SSW-SLH003, QYZDJ-SSW-SLH040; 100 Talents Program of CAS; The Institute of Nuclear and Particle Physics (INPAC) and Shanghai Key Laboratory for Particle Physics and Cosmology; ERC under Contract No. 758462; European Union's Horizon 2020 research and innovation programme under Marie Sklodowska-Curie grant agreement under Contract No. 894790; German Research Foundation DFG under Contracts Nos. 443159800, 455635585, Collaborative Research Center CRC 1044, FOR5327, GRK 2149; Istituto Nazionale di Fisica Nucleare, Italy; Ministry of Development of Turkey under Contract No. DPT2006K-120470; National Research Foundation of Korea under Contract No. NRF-2022R1A2C1092335; National Science and Technology fund; National Science Research and Innovation Fund (NSRF) via the Program Management Unit for Human Resources \& Institutional Development, Research and Innovation under Contract No. B16F640076; Polish National Science Centre under Contract No. 2019/35/O/ST2/02907; Suranaree University of Technology (SUT), Thailand Science Research and Innovation (TSRI), and National Science Research and Innovation Fund (NSRF) under Contract No. 160355; The Royal Society, UK under Contract No. DH160214; The Swedish Research Council; U. S. Department of Energy under Contract No. DE-FG02-05ER41374.
\end{acknowledgments}


%

\end{document}